%% file: main.tex
\documentclass{sig-alternate}
\usepackage{url} 
\makeatletter
\g@addto@macro{\UrlBreaks}{\UrlOrds}
\makeatother

\usepackage{flushend} 
\usepackage{graphicx} 
\usepackage{subcaption} 
\usepackage{epstopdf} 
\usepackage{amsmath} 
\usepackage{algorithm}
\usepackage{algpseudocode}
\usepackage{caption} 
\newcommand{\ignore}[1]{}

\usepackage{leading}

\begin{document}

\title{Vulnerable GPU Memory Management: Towards Recovering Raw Data from GPU}

\numberofauthors{3}
\author{
\alignauthor
Zhe Zhou, Wenrui Diao, Xiangyu Liu\\
       \affaddr{The Chinese University of Hong Kong}\\
       \email{\{zz113, wr013, lx012\}@ie.cuhk.edu.hk }
\alignauthor
Zhou Li\\
       \affaddr{ACM Member}\\
       \email{lzcarl@gmail.com}
\alignauthor
Kehuan Zhang, Rui Liu\\
       \affaddr{The Chinese University of Hong Kong}\\
       \email{\{khzhang, ruiliu\}@ie.cuhk.edu.hk }
}
\maketitle


\begin{abstract}


In this paper, we present that security threats coming with existing GPU memory management strategy are overlooked, which opens a back door for adversaries to freely break the memory isolation: they enable adversaries without any privilege in a computer to recover the raw memory data left by previous processes directly. More importantly, such attacks can work on not only normal multi-user operating systems, but also cloud computing platforms.

To demonstrate the seriousness of such attacks, we recovered original data directly from GPU memory residues left by exited commodity applications, including Google Chrome, Adobe Reader, GIMP, Matlab. The results show that, because of the vulnerable memory management strategy, commodity applications in our experiments are all affected.

\end{abstract}



\ignore{It may leaks user's sensitive data if the isolation is not solid enough. The prevalence of clouding computing with GPU even exacerbated the consequence. Malicious machine may reside in the same physical machine as yours and may steal your sensitive GPU memory without your awareness and permission by exploiting the weak isolation. This paper presents how adversary can recover user's image by exploiting the leakage of non-zeroed video memory.}
\input{intro.tex}
\input{background.tex}

\input{adv.tex}
\input{image.tex}

\input{evaluate.tex}
\input{end.tex}

{ \bibliographystyle{abbrv}
\bibliography{main}}



\end{document}

%% file: intro.tex
\section{Introduction}
Graphics Processing Unit (GPU) has become an indispensable component in today's computing systems. To efficiently handle graphic processing tasks, its architecture can support highly parallel computations. This distinctive feature also extends its capabilities beyond graphic processing: along with the emerging of GPGPU (General-Purpose Computing on GPU) technique, GPU is utilized for a broad spectrum of computing tasks, like genome sequencing, signal processing, etc.

Unfortunately, the boost in performance sacrifices security. For efficiency, discrete GPUs are equipped with exclusively used and heterogeneous memory system, which is managed by GPU independently and is out of the control of CPU, the center of the computing system. Such design introduces the potential of memory isolation issue: isolation policies enforced by operating system and executed by CPU may not identically performed.

The security issue regarding GPU memory management was overlooked, because the undocumented and close-source design memory system hindered people's way to realize its seriousness. No corresponding patch has been released by the mainstream GPU vendors since the first time people cast doubt on the security of the strategy~\cite{leakageweb}. The reported security potention was labeled as low-risk, because the data stored in the GPU memory cannot be directly recovered due to the undocumented structure. Even in the latest study, side-channel attacks over the GPU disclosed users' browsing history, but still didn't expose the seriousness.

In this paper we investigate this problem in-depth and argue that \textbf{the security risks of contemporary GPU memory architectures are underestimated.}  To this end, we examine possible attacks that can recover raw data from GPU memory residues left behind by innocent applications and extract sensitive information within. Our study and evaluation results show that adversaries are indeed able to get original images, texts and matrix data that are all highly sensitive.


The path to successful attack is however obscured. We have to overcome several challenges, especially in image recovery. We need to identify tiny image-like objects (varies from several kilobytes to several megabytes) and their formats from a big memory space (usually in several gigabytes). We also need to infer image's layout, i.e. the length and width. To point out, such information is not preserved in GPU memory. We tackle the first challenge through exploiting the unique combinations of the byte values of image pixels. For the second challenge, we design a novel algorithm based on one key insight on image data: when examined in frequency domains, the similarities between adjacent rows of an image should show some cyclical pattern, and in ideal situations, the cycle is equal to image width.  With these techniques, the boundary and format of an image can be determined and the image can be recovered without quality loss.

\ignore{
the inherent feature of a meaningful image - that the consecutive rows and columns are correlated and should be similar. \ignore{In particular, our approach investigates the spectrum of the memory block and infers the dimensional information according to the periodicity of the spectrum after a series of transformations.} In particular, we look into the spectrum of the image matrix and infer the layout based on the periodicity of spectrum. With this  information, the boundary and format of an image can be determined and the image can be recovered without quality loss.
}

\begin{figure}
\centering
\begin{subfigure}[b]{0.5\textwidth}
    \center
    \includegraphics[width=0.6\textwidth]{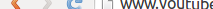}
    \caption{Address Bar}
    \label{Addr}
\end{subfigure}

\begin{subfigure}[b]{0.5\textwidth}
    \center
    \includegraphics[width=0.6\textwidth]{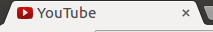}
    \caption{Tab Caption}
    \label{Tab}
\end{subfigure}
\caption{Recovered image From GPU memory indicating the URL of the displayed web page}
\end{figure}

The consequences of the vulnerable memory management strategy presented in this paper are much more serious than previous researchers expected. \textbf{First}, the raw data recovered from memory is \textbf{highly sensitive}. While previous researchers thought GPUs have some undocumented mechanism to hide the plain text memory, such that only side-channel information like distribution statistics were used to infer what was processed on GPU~\cite{LeeKKK14}. \textbf{Second}, our method can work with a wide spectrum of commodity GPU-accelerated applications. We selected some popular GPU-accelerated applications with large user base to test our attack, including Google Chrome (browser), Adobe Reader (document processor), GIMP (image processing), Matlab (scientific computing). All of them are vulnerable in the end. For example, Google Chrome employs GPU to perform high speed page rendering, and we are able to reconstruct the image fragments within address bar, tab and page content. Examples of recovered address bar and tab are shown in Figure~\ref{Addr} and Figure~\ref{Tab} and the web site address can be easily recognized from them. What's more, we show that \textbf{account name} and even \textbf{email titles} of a user can also be obtained (see Section~\ref{chrome}). Another interesting case is Adobe Reader, in which we show the text fragment embedded within PDF document can be recovered. Considering nowadays GPU acceleration is increasingly adopted by mainstream software (e.g., Microsoft Office and Libre Office), it is expected that more and more sensitive data would be sent to GPU for processing, and then exposed to adversaries if the attacks are launched. \textbf{Third}, besides the traditional multi-user systems, attacker can also launch attacks on the virtualized platform (e.g., in cloud scenario), which further extends the victim population to those well protected systems adversaries could not access directly.


\textbf{Contributions.}
We summarize this paper's contributions as follows:

\begin{itemize}

    \item \textit{We firstly identified that GPU memory management strategy is vulnerable.} We found that the GPU memory management strategy can be exploited by malicious programs to cross the memory isolation boundary to get the raw memory data belonged to other processes, which is highly risky and leads to much more serious consequences than researchers expected before.

  \item \textit{A novel methodolgy to recover original images from GPU memory residues.} We proposed a new approach that can automatically identify and recover images from the GPU memory residues left behind by legitimate applications. Based on our insights on image data and signal processing techniques, we designed a new algorithm that can determine image layout and extract images effectively. Compared with previous works on GPU security issues, our approach is the first to show that original images and sensitive information can be directly recovered from GPU memory residues.

  \item \textit{In-depth evaluation.} We evaluated our attacks against popular applications with each coming from one typical use of GPUs, including Web browsing, document processing, photo editing, scientific computing etc. The results show that all of them are vulnerable to such attacks. Besides the ordinary multi-user systems, we further tested our approach on virtualized platforms and find it also works. 



\end{itemize}

\textbf{Roadmap.}
The rest of the paper is organized as follows. Section~\ref{backgnd} presents background knowledge around GPU architecture and its security implications. Section~\ref{model} describes the adversary model and our assumptions. Section~\ref{algo} describes how to get memory residues and how the algorithm recovers image from them. Section~\ref{eval} elaborates the test settings and evaluation results. \ignore{Section~\ref{defense} points out out limitations of our work.} Section~\ref{relat} summarizes related works, followed by Section~\ref{concl} that concludes the paper.

%% file: background.tex
\section{Background}
\label{backgnd}
In this section, we first overview the computing model of GPU. Then the security issue about the memory management is reviewed in the end and we highlight the contributions of our attack.

\subsection{GPU Computing Model}
GPU is responsible for highly parallel computing works, which resulted to a very different architecture from CPU. GPU has a large amount of computing units and its independent memory chip. To compute, data must be copied from main memory that is controlled by CPU and OS to GPU memory that is invisible to CPU. GPU manages and operates its own memory. After GPU finished the computing, the result is copied back to main memory.

Users can only operate GPU as well as the GPU memory through APIs provided by GPU like OpenCL or OpenGL. OpenCL APIs already allow users to operate memory nearly natively.
\subsection{Vulnerabilities in GPU}

GPU is designed for the purpose of more efficient computing, but the security implications are not fully studied. The GPU manages its own memory, neglecting the policy enforced by operating system, which resulted in an inconsistency between two memory management strategies. Main memory managed by CPU is not cleared immediately when it is freed, but operating system guarantees that the memory read by a process without any initialization is zero. GPU however does not provide such guarantees. Previous works demonstrated that the the memory read out without initialization is not zero.

Without documentation to the memory management strategy, people did not recover data directly from the memory, but it does not stop researchers from launching side-channel attacks. For example, Lee in \cite{LeeKKK14} demonstrated that the memory has different bytes distribution when user visit different web pages, so it can be inferred that which sites are visited by the user, though the bar of the attacks is high: attackers must profile a pile of web pages to infer which site is visited by the user~\cite{LeeKKK14}. And the granularity of the information recovered is coarse: only which one among a set can be inferred, so it cannot be inferred if the user visits unpopular web sites. 

\subsection{GPU in Virtualized environments}
Nowadays, all mainstream cloud platforms provide GPU equipped virtual machines~\cite{cloudGPU}. Different from other resources like network and storage which can be easily shared by different VMs, a GPU is often assigned to a dedicated VM in virtualized environment~\cite{maurice2014confidentiality}. The technique assigning a GPU to a VM is named \textit{GPU Passthrough}.

With a passed through GPU, computation using GPU on VM can reach bare-metal level. After the VM finishes its computation and shuts down, the GPU of hypervisor is kept powered up because physical machine is not shutdown. Then hypervisor can assign the GPU card to another VM depending on scheduling.

%% file: adv.tex
\section{Adversary Model}
\label{model}

The adversary model in this paper is the same as what has been specified by previous works \cite{maurice2014confidentiality,LeeKKK14,di2013cuda}. The targeted machine is equipped with discrete GPU which supports computing APIs (i.e., OpenCL or CUDA). We assume that an adversary has successfully acquired the permission to run malicious code under an \textbf{unprivileged account} on the target machine. \ignore{However, the adversary could not steal any information directly from the victim user's account because the victim's account has not been compromised and is well-protected by underlying operating system.} What the adversary wants to do is to bypass memory protection and get sensitive information left by other processes by only reading and analyzing the GPU memory without special privilege.

In cloud setting, we assume that an adversary can rent a GPU equipped virtual machine from service provider. Victim users also possess a virtual machine on the same physical machine. Once the victim shutdowns his VM and the adversary starts his VM, adversary can dump the GPU memory left by the VM of victim, because GPU does not lose power and does keep all the data on memory during the VM switching.

\ignore{Such scenarios are in fact very common. For example, computers of lab or library in most universities are shared among students and staffs. In some companies or factories, a shared computer may be used for high-profile tasks such as processing secret documents or accessing an isolated network. An attacker could login and start a malicious background program which silently collects sensitive data from other users whoever later login and use the same computer.  \ignore{Attackers could exploit such configurations to steal information from other users even with an unprivileged account.}}


The adversary only needs the capability to access (read and write) the GPU memory, which does not require any restricted permission. By writing the GPU memory, the adversary could mark out memory regions not possessed by applications, and by reading the GPU memory, the adversary could dump current GPU memory or examine the status of GPU memory allocation (e.g., to identify the sudden and large increase of available memory indicating that the victim application just deallocated a large chunk of used memory~\cite{LeeKKK14}).

%% file: image.tex
\section{Image Recovering}
\label{algo}
In this section, we propose a method to recover graphical data from the GPU. The most important task of GPU is graphic processing, so if attacker can recover original graphic data from the GPU, it implies that the GPU memory management strategy is highly vulnerable. We first present techniques about how to identify image-like tile from GPU memory. The boundary identified in the step is not precise. Then, we describe how to reconstruct image from the tile, i.e., how to infer the image layout including image width and length. At last, the paper shows how to precisely rearrange the recovered image.

\subsection{Tile Extraction}
\label{acquisition}

When an application utilizes GPU to accelerate image rendering, the image content will be loaded into GPU memory at some point. Our initial step is to extract the data blocks (or \textit{tiles}) which are likely to contain or be part of meaningful images. This turns out to be a non-trivial task for two reasons. First, the metadata about the image objects is not stored in GPU memory. In other words, the location and layout of the images objects are unknown to adversary. Second, GPU is also used for general-purpose computations like encryption and non-graphical blocks might be left in memory and mixed with image objects. We leverage several distinctive features of images to identify the tiles. We modified the prime-probe method used in \cite{LeeKKK14} to extract memory and the process is elaborated below:

\vspace{2pt}\noindent{\bf Memory initialization.} Our attack targets images left by applications of interests (like browser) and the data generated in other cases are not considered. The memory regions which are used to keep such images are recognized during the memory initialization procedure. Before the start of the victim applications, the malicious program marks the whole video memory with $0xff$ (e.g., 512 MB for AMD Radeon HD 6350) using GPGPU API (e.g., \texttt{clCreateBuffer} or \texttt{clEnqueueWriteBuffer} in OpenCL). Thus, the data left by applications terminated ahead are all wiped out and the newly allocated data can be attributed to the targeted running applications.

\ignore{If there are other applications using GPU in this stage, their occupied memory will not be flushed. In this instance, it could happen that the memory size requested exceeds the size of available GPU memory. However, the memory sharing mechanisms like Hypermemory and Turbocache will automatically use space from main memory to satisfy the need for extra memory, which avoids exceptions.}

\vspace{2pt}\noindent{\bf Data blocks extraction.} After the memory is initialized, the malicious program runs at the background and queries for the size of available memory periodically. If the size of available memory increase, a chunk of memory is deallocated by the running applications, and the malicious program will collect the memory residues. Here, we need to check if the applications of interests are running to avoid unnecessary analysis, i.e., collect memory used by irrelevant applications. The list of running processes is queried at the same time (e.g., we use \textit{ps} command to read process list on Linux platform) and the analysis is continued only if the targeted applications are within the list. We again use GPGPU API (e.g., \texttt{clEnqueueReadBuffer} in OpenCL) to extract meaningful data blocks. Since GPU processor does not clear the memory residues of applications and developers in most cases do not wipe out the used memory, there is high chance that sensitive information is remained in the dump. \ignore{Occasionally, the sensitive information is overwritten by another application prior to be read by the malicious program. The probability is nonetheless small given that a legitimate application has no incentive to use the memory space just released by another application.}

One may think that the images can be easily extracted from memory dump by removing all $0xff$ bytes. Unfortunately, this simple approach is ineffective as $0xff$ is also legally used to represent pixel's color and alpha component value, so if naively remove all $oxff$, graphical data will be broken down.

\ignore{On the other hand, the pixels of image are stored sequentially in GPU memory, because the computing API does not provide 2D array support. In fact,  In this way, the image is very likely to be sequentially stored to GPU memory.}

This motivates us to identify blocks constituting the image and then stitching them together. Our strategy is to divide the dumps into blocks of fixed size, and merge the ones that are consecutive and used by applications. The block size has to be determined first. The size should be smaller than a image because otherwise two or more images would be in the same block and be regarded as one image. The block size can neither be too small because otherwise big white chunk (all $oxff$ trunk) in a image would break image down. After a lot of experiments, we chose 4K as the block size that works well in most cases. Therefore, we split the memory into 4K-size blocks and filter out the blocks that are filled with $0xff$ as they are probably not used after initialization. We also remove the blocks that are all $0x00$ since they are clean blocks zeroed out by developers or OS. The blocks are \textbf{concatenated into a bigger block} if they are consecutive in memory space, and we call it tile. After this step, the data blocks left by victim processes are extracted.

The structure of the graphical data is unknown. As mentioned before, GPU manufacture didn't provide documentation about how they map the logic address into physical address. But, at least, we are convinced that developer will not disorder the memory of an image in the logic address space, because computing API does not provide 2D array support so developers often store 2D objects like image in GPU memory using 1-D vector.\footnote{We search the term "opencl 2D array" in Google, and the top results all advise readers to use the code statement like ``\#define A(x,y) a[x*width + y]'' to emulate 2D array.}

We found GPUs do have a lot of storage techniques to achieve higher memory performance. For example, according to the advertising document, recent GPUs have memory compression which automatically compresses the data to be stored into the memory using delta algorithm and decompresses the memory automatically when it is accessed. There raises a question that whether those techniques will disorder the pixels in the tiles. After reviewed those techniques, we found that they are all transparent to upper layer, which means that no matter how data are re-ordered, compressed and encoded, the data read out by program are the original ones. According to our evaluation result, the byte sequences in data blocks are retained.

\vspace{2pt}\noindent{\bf Data blocks pruning.} The obtained blocks in the last step still need further pruning - blocks might be used to keep non-graphical data and they are not considered by us for now. Favorably, the distinctive structure of image's data helps us to identify the graphical blocks. For one image, each pixel is represented by a 4-byte word, or 4 8-bit \textit{channel}. The 4 channels correspond to color Red (R), Green (G), Blue (B) and Alpha component (A) value of the pixel separately. The alpha component value indicates the level of transparency of the pixel. From the survey over a large number images, we found that this value is either $0x00$ or $0xff$, indicating the pixel is mostly set to be nontransparent or the channel is unused. So, we can judge if the data block is graphical by checking its alpha channel values.

For a graphical data block, we also need to determine the order of channels to guide the later reconstruction step. In theory, developers can choose any order but they usually use the first or last byte of the 4-byte word for alpha channel and sequentially align RGB values for compatibility. The common image format is therefore either RGBA or ARGB. To find out which format is used, we compute the percentage of $0xff$ or $0x00$ stored in each byte of 4-byte word ($p$), and compare it against a threshold ($th$). If $p>th$ for the last byte, the image format is considered as RGBA. If $p>th$ for the first byte, the image format is considered as ARGB. Otherwise, the block is discarded. In the evaluation, $th$ is set to 20\%.


At last, we remove the heading and trailing elements filled with values of $0x00$ or $0xff$ and only keep the part in the middle (It does not mean that the boundary is now precisely determined). Ideally, a tile contains and only contains one image and this has been proved to the dominant case by our pilot experiments. However, we did observe a small portion of tiles which contain parts from two or more images, because the locations of the included images in the memory space are too close. \ignore{ The boundary between two different images can be identified by seeking the block of consecutive $0x00$ or $0xff$ bytes within the tile. Thus, we search for the region with more than $th'$ $0x00$ or $0xff$ bytes and $th'$ is set to be 2K. If such region is identified, we remove it and split the tile into 2 sub-tiles. This process is applied recursively on the sub-tiles till no blank sub-block is identified. We describe our approach of reconstructing images from tiles next.}


\input{inference}

%% file: inference.tex
\subsection{Image Layout Inference: Problem}
\label{problem}

After a tile is extracted, the next step is to infer the layout information associated with the embedded image. Assume the tile occupies $N$ 4-byte words in memory and the image occupies $W$ 4-byte words ($W \le N$) sequentially. The image could reside at any sub-area of the tile. We denote the number of 4-byte words ahead of the image as $s$ and the number after as $e$ and $N=s+W+e$. We need to identify $s$ and $e$ to retrieve the sub-area. Since an image is represented with a 2-dimensional matrix, we also need to identify the number of rows (say, $n$) and columns (say, $m$, which equals to $W/n$) to recover the original image.


\begin{figure}[h]
    \center
     \begin{subfigure}[h]{0.22\textwidth}
        \includegraphics[width=\textwidth]{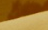}
        \caption{Normal Image}
        \label{fig:similar}
     \end{subfigure}
     \begin{subfigure}[h]{0.22\textwidth}
        \includegraphics[width=\textwidth]{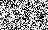}
        \caption{Signal-less TV Image}
        \label{fig:tv}
     \end{subfigure}
     \caption{Normal Image and Random Image}
\end{figure}

Commonly, the size of an image ranges from several KB to several MB. It is infeasible to enumerate all the combinations of $s$, $e$ and $n$, and then let the attacker to decide which combination can lead to the restoration of the original image. On the other hand, an image (especially the sensitive one) is quite different from other artificially generated data: \textbf {there lies strong similarity between consecutive rows and consecutive columns} (see Figure~\ref{fig:similar}). Another favorable condition is that though an image could be compressed when stored at hard disk or transmitted through network, it is decompressed and usually loaded into matrix structure in GPU memory and the similarity is preserved. Transparent memory acceleration techniques may not break the similarities because hardware guaranteed that upper layer application will not observe their existence except performance increases. We leverage this key insight to infer $n$ or $m$ and henceforth $s$ and $e$ (the details are described in Section.\ref{approach}). Our approach, however, is not designed to recover randomly generated image like the screen of analogy television when there is no signal (see Figure~\ref{fig:tv}) or an image filled with identical pixel. These types of images usually do not enclose sensitive information and are disposed.


\subsection{Image Layout Inference: Approach}
\label{approach}
When processed by GPU, an image is usually stored in a 2-dimensional matrix (denoted by $a$), and the value of a pixel can be read from $a[i,j]$, where $i$ and $j$ denotes the $i_{th}$ row and $j_{th}$ column in the image matrix. On the other hand, a tile is just a sequential data block represented by an 1-dimensional vector (denoted $f$) while $a[i,j] = f[s + i \times m + j]$. Our goal is to infer the correct $s$ and $m$ which fulfills this equation. As stated in Section~\ref{problem}, the consecutive rows of an image are similar ($a[i,:]$ is similar to $a[i+1,:]$, for $0 \le i \le n-2$), and we leverage this constraint to find the correct $s$ and $m$.\ignore{this observation inspired us that the distinctive similarity may imply the value $s$ and $m$ in some way. In other words, some parameters related to similarity may be correlational to $m$ and $s$. So $m$ and $s$ could be calculated given the similarity related parameters.}

\ignore{After trying a lot of methods, we found that the amplitude spectrum of the image reflects the periodicity of the similarities. That's to say $m$ can be calculated after having the spectrum. Besides, $s$ doesn't interfere the calculation of $m$ in this stage with some processes, so $s$ can be handled later after $m$ is correctly calculated.} However, we still need an appropriate metric to quantify similarity between rows. By examining different metrics, we found out the best one is the amplitude spectrum in the frequency domain of image matrix. If $m$ is correctly inferred, the distribution of element values in each row should be similar, leading to strong periodicity of row values. In the subsequent paragraphs, we introduce an algorithm which first infers $m$ and then derives $s$ based on $m$. Our approach is demonstrated through four types of tiles with increasing difficulty for processing. For each type, we remove one constraint from the previous type and the final type reflects the tile extracted from genuine GPU memory dump. In the end, we solve the number of redundant 4-byte words ahead and behind the image ($s$ and $e$). Throughout this section, we use tiles shown in Figure 3 as examples to motivate our approach.

\begin{figure*}
    \begin{subfigure}[ht]{0.5\textwidth}
        \centering
        \begin{subfigure}[h]{0.45\textwidth}
            \center
            \includegraphics[width=\textwidth]{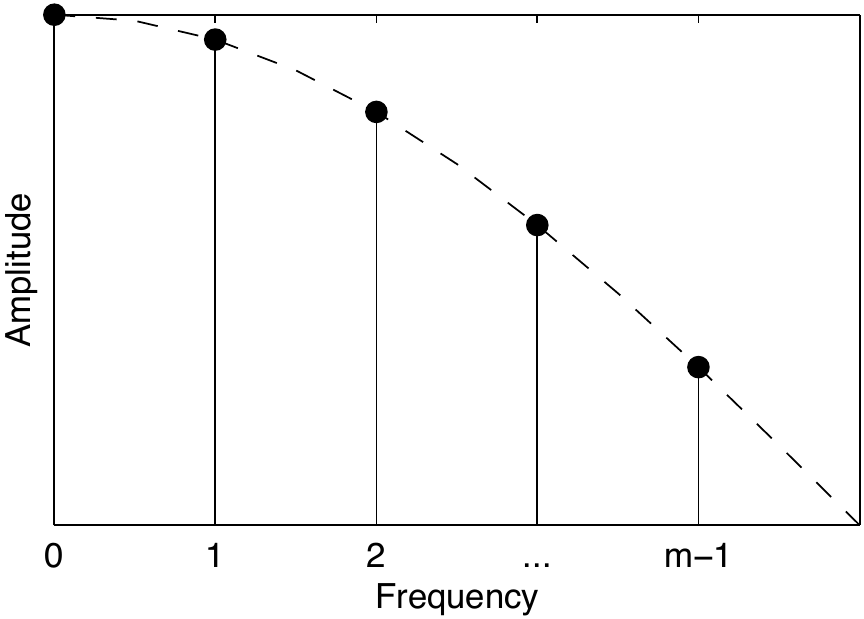}
            \caption{}
            \label{specr}
        \end{subfigure}
        \begin{subfigure}[h]{0.45\textwidth}
            \center
            \includegraphics[width=\textwidth]{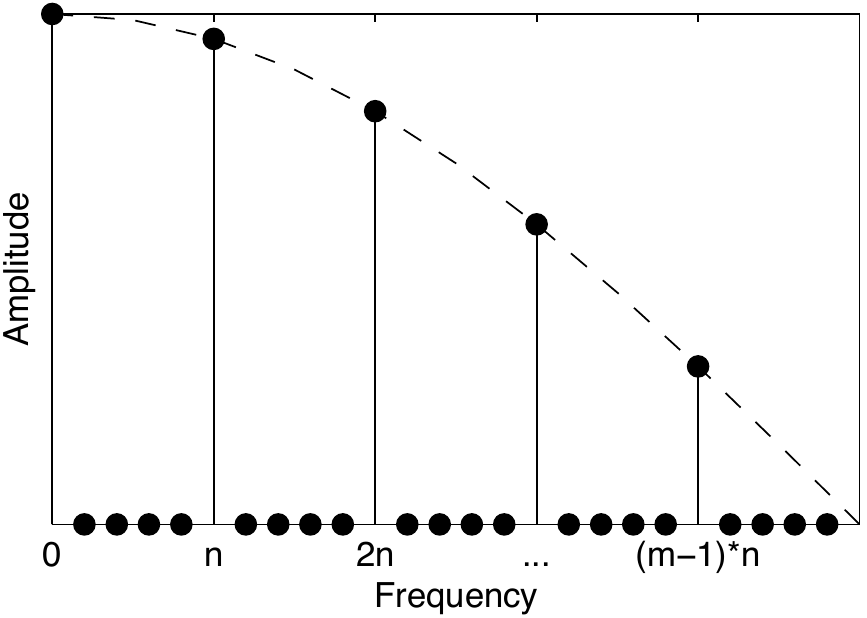}
            \caption{}
            \label{speci}
        \end{subfigure}
        \begin{subfigure}[h]{0.45\textwidth}
            \center
            \includegraphics[width=\textwidth]{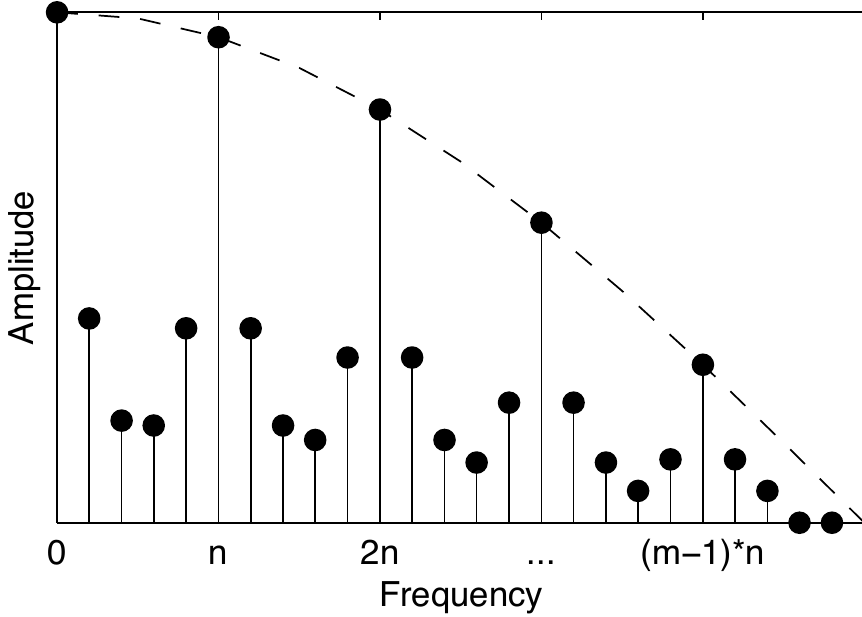}
            \caption{}
            \label{spec3}
        \end{subfigure}
        \begin{subfigure}[h]{0.45\textwidth}
            \center
            \includegraphics[width=\textwidth]{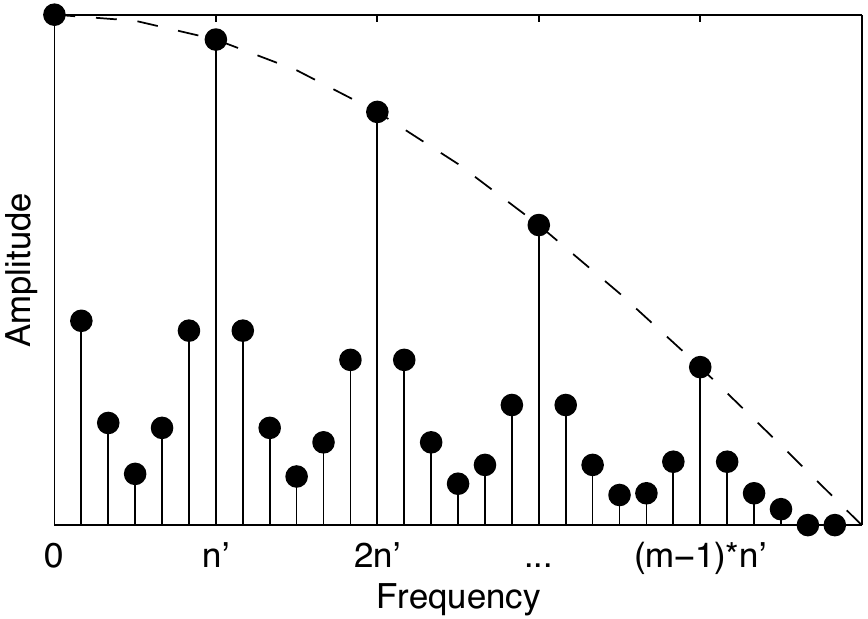}
            \caption{}
            \label{spec4}
        \end{subfigure}
    \end{subfigure}
    \begin{subfigure}[ht]{0.28\textwidth}

         \begin{subfigure}[h]{\textwidth}
            \includegraphics[height=0.025\textheight]{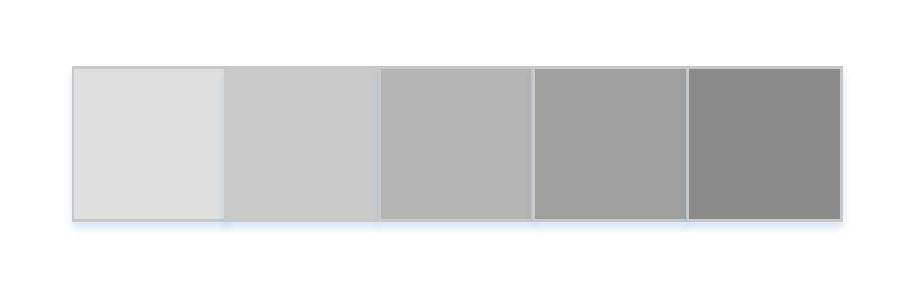}
            \caption{}
            \label{row}
         \end{subfigure}

         \begin{subfigure}[h]{\textwidth}
            \includegraphics[height=0.025\textheight]{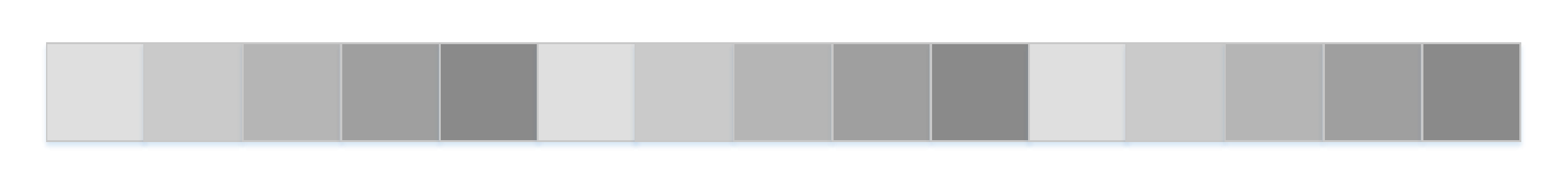}
            \caption{}
            \label{cond1}
         \end{subfigure}

         \begin{subfigure}[h]{\textwidth}
            \includegraphics[height=0.025\textheight]{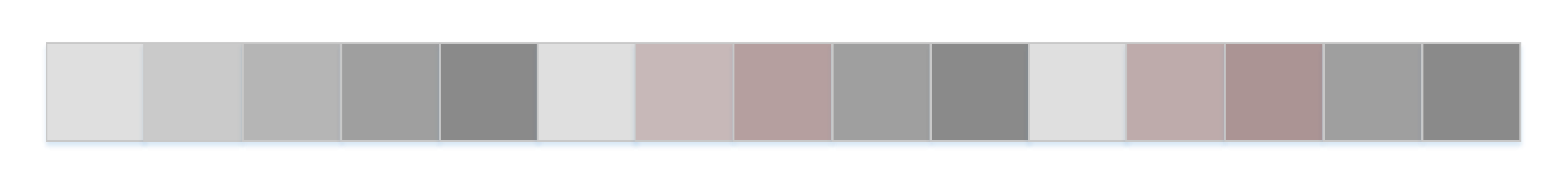}
            \caption{}
            \label{cond2}
         \end{subfigure}

         \begin{subfigure}[h]{\textwidth}
            \includegraphics[height=0.025\textheight]{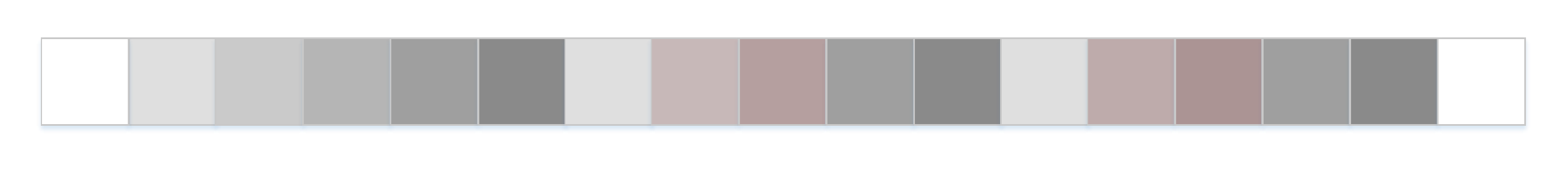}
            \caption{}
            \label{cond3_1}
         \end{subfigure}

         \begin{subfigure}[h]{\textwidth}
            \includegraphics[height=0.025\textheight]{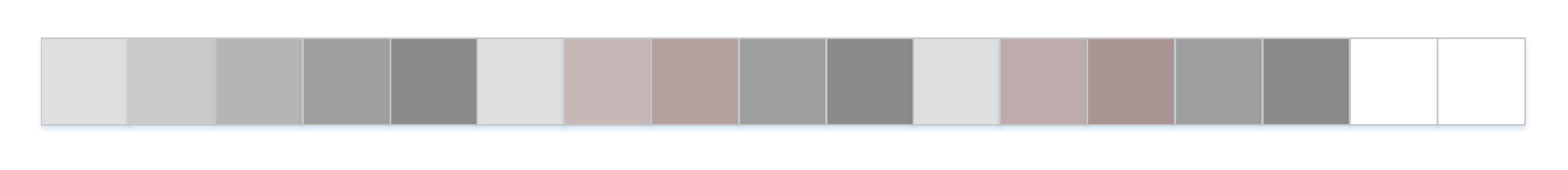}
            \caption{}
            \label{cond3_2}
         \end{subfigure}



    \end{subfigure}\quad\quad
    \begin{subfigure}[h]{0.15\textwidth}
         \begin{subfigure}[h]{\textwidth}
            \includegraphics[width=\textwidth]{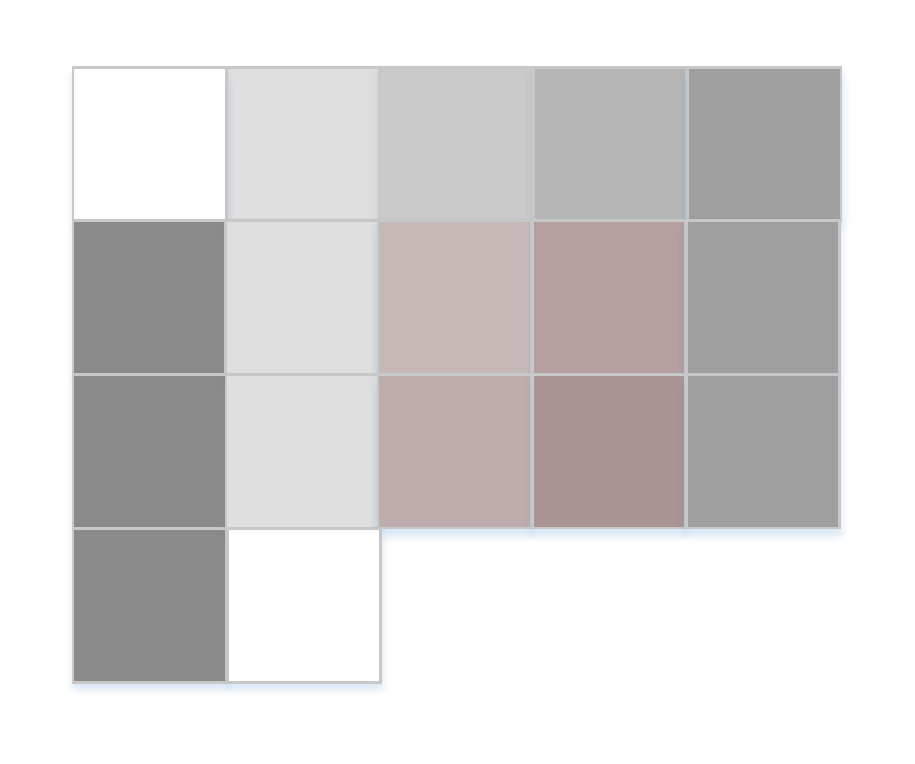}
            \caption{}
            \label{all_1}
         \end{subfigure}
         \begin{subfigure}[h]{\textwidth}
            \includegraphics[width=\textwidth]{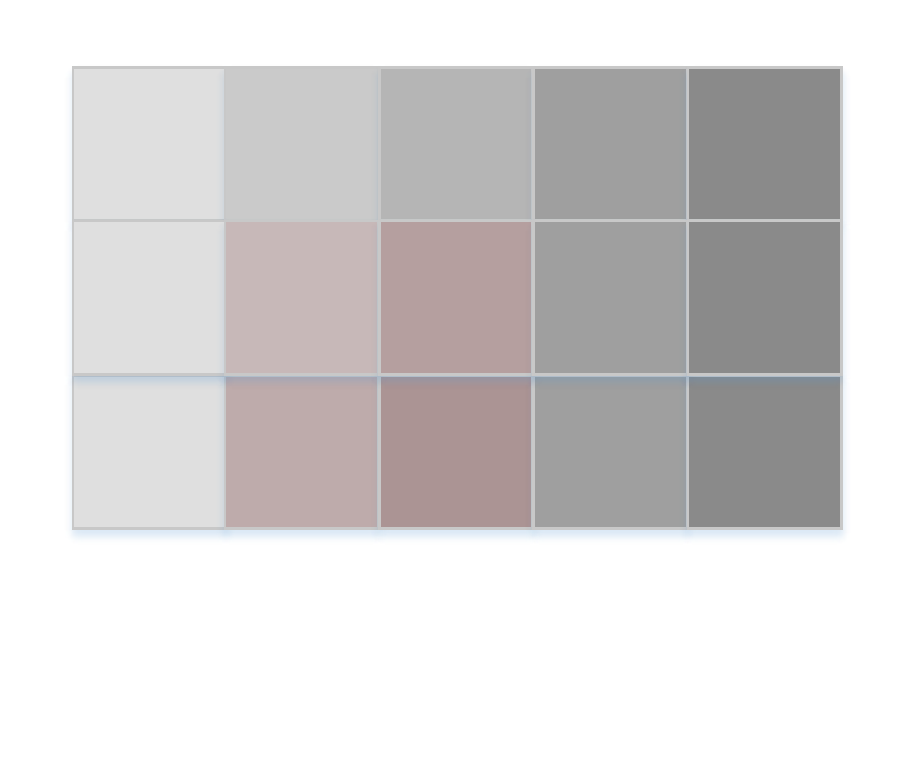}
            \caption{}
            \label{all_2}
        \end{subfigure}
    \end{subfigure}
    \caption{Images used to illustrate 3 types of images}
\end{figure*}

\vspace{2pt}\noindent{\bf Tile type I}
We start from an easy case where the similarity can be trivially quantified. We assume there is no trailing and leading redundant pixels and all rows are identical ($s=0$, $e=0$ and $\forall i_1,i_2 \in [0,n-1], a[i_1,:] = a[i_2,:]$). \ignore{The pixels within each row could be arbitrary but the neighboring columns should be similar.} We illustrate such type of tile in Figure~\ref{cond1} in which one row filled with distinct pixels (see Figure~\ref{row}) is duplicated for three times. In this case, $f$ is turned into a periodic function where $f[x] = f[x+m], \forall x \in [0, (n-1) \times m - 1]$ and $m$ is the interval. Here, we leverage spectrum produced by \textbf{FFT (Fast Fourier Transform)} algorithm to capture this interval. Fourier Transform can decompose a signal from time domain into frequency domain and is widely used in signal processing and image processing, etc\cite{FFT,DFT}. 

\ignore{
\begin{figure}
 \begin{subfigure}[h]{0.15\textwidth}
            \includegraphics[width=\textwidth]{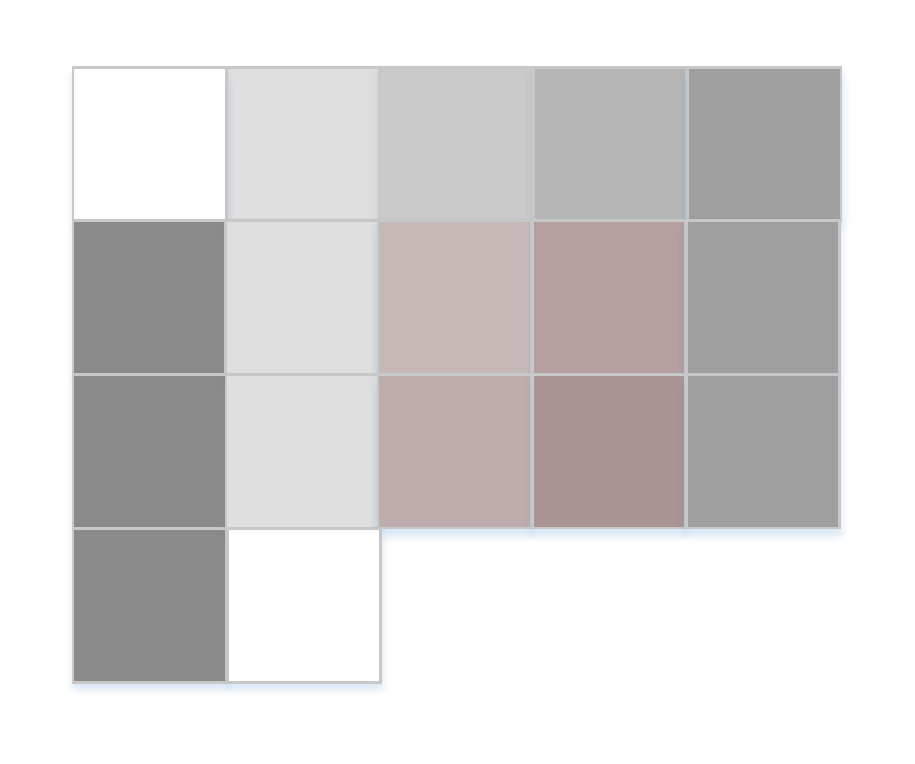}
            \caption{Reshaped block using the calculated $m$}
            \label{all_1}
         \end{subfigure}
         \quad \quad 
         \begin{subfigure}[h]{0.18\textwidth}
            \includegraphics[width=\textwidth]{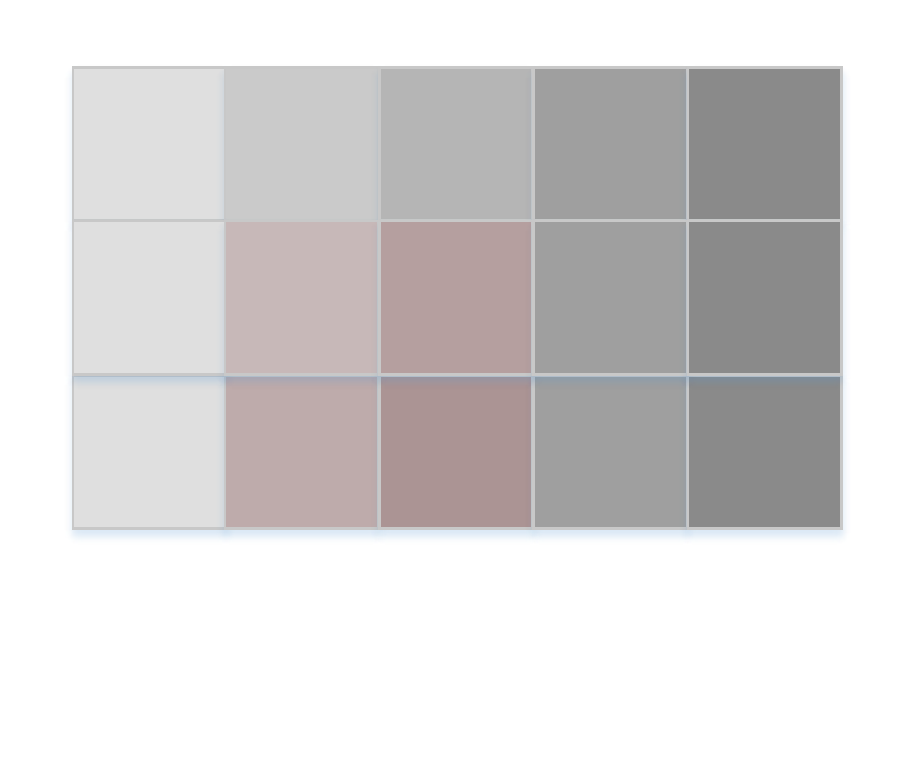}
            \caption{Reshaped and post-processed block}
            \label{all_2}
 \end{subfigure}
\end{figure}
}
We denote \textit{amplitude spectrum} of $f$ produced by FFT by $F$ (pixels are gray-scaled before FFT). We studied $F$ and found out that the interval between two non-zero components equals to $n$, the number of rows. So, in this case, the image can be easily recovered, we demonstrate the prove for this observation below:

$F$ for this tile is illustrated in Figure~\ref{speci} and Figure~\ref{row} shows the spectrum of only a row $F_0(k)$. $F(kn) = F_0(k)$ for $k=0, 1, 2, ..., (m-1)$) and $F(kn)$ are non-zero (we call them main components) while the the amplitudes for other components are zero, according to the properties of periodical signal. So, the interval between two neighboring main components equals to the number of rows ($n$) of the image matrix. The width $m$ can be computed through $N/n$ and the image is therefore recovered by reshaping the tile using those parameters. With the parameters, image can be recovered.


\vspace{2pt}\noindent{\bf Tile type II.}
Images we encountered normally do not have such a property that all rows are identical. In this case, we assume that neighboring rows are similar but not always identical. Still, we assume there is no element in the tile ahead or end. We found that again FFT can be used to infer the number of rows and columns of the image matrix.

As an example, we assume the tile looks like Figure~\ref{cond2} and the amplitude spectrum $F$ of the tile vector $f$ is illustrated in Figure~\ref{spec3}. This time, the main components of $F(k)$ occur when $k=0, n, 2n, ..., (m-1) \times n$, which is the same as the spectrum of tile type I. However, due to the differences between two neighboring rows, the main components disperse and the amplitudes of the non-main components are greater than zero now. Still, they are much smaller than those of main components and the main components can be easily identified. We explain the scenario below:

We introduce $n$ virtual images $v_1, ..., v_n$ and all of them have the same layout as the original image. Particularly, $v_i$ is constructed through replicating the $i_{th}$ row of $a$ for $n$ times, thus $\forall i, x \in [0,n-1], v_i(x,:) = a[i,:]$.
We denote the amplitude of $v_i$ for the $k_{th}$ element by $V_i(k)$ and obviously it equals to $F_i(\frac{k}{n})$ according to the previous analysis, if $k$ is a multiple of $n$. For a sample tile shown in Figure~\ref{cond2}, the value of an element can be represented by $f(i \times n + j)$ and also by $F_i(k)$ through inverse FFT:

\begin{equation*}
    \begin{split}
        f(i \times n + j) = v_i(x,j)&=\frac{1}{N}\sum\limits_{k=0}^{N-1}W_N^{-k(xm+j)}V_i(k)\\
        &=\frac{1}{N}\sum\limits_{k=0}^{N-1}W_N^{-k(xm+j)}F_i(\frac{k}{n})\\
        &=\frac{1}{N}\sum\limits_{k=0}^{m-1}W_N^{-knj}F_i(k)\\
        where \;W_N\; is\; the\; & twiddle \;factor\\
    \end{split}
\end{equation*}

The above equation suggests that $f$ consists of sub-components with frequency of the multiple of $n$. Based on our observation that the consecutive rows vary slightly, the differences between $F_i(k)$ and $F_{i+1}(k)$ should also be small, and the combined $F(k)$ should be large when $k$ is the multiply of $n$. For other $k$, the combined $F(k)$ is still small, which makes the main components stand out at $0, n, 2n, ..., (m-1) \times n$. Similarly, we compute the interval between two main components to derive the value of $n$ and then $m$. \ignore{which suggests that each main-component has only a narrow dispersion width. The spectrum of frequency points except $nk$ are no longer zero but much smaller comparing to $nk$ points. Hence, a component is considered as a main one if its spectrum is much higher than the neighboring ones.} 

\vspace{2pt}\noindent{\bf Tile type III.}
Next, we consider the case that there are a block of pixels ahead of and another block of pixels trailing the original image object and the value of each element in the blocks is zero. As stated in the theorem in DFT~\cite{DFT}, the amplitude spectrum does not change when circularly shifting the original signal. So the leading block ahead of the image, if any, can be shifted to its end without any impact to the spectrum. We illustrate the image with both leading and trailing block in Figure~\ref{cond3_1} and the image with only trailing block in Figure~\ref{cond3_2} and their spectrum are the same (see Figure~\ref{spec4}). We also assume the trailing block is filled with zero to avoid the disturbance of non-zero padding to the original spectrum in this simplified condition.

\begin{figure}[ht]
\centering
\includegraphics[width=0.45\textwidth]{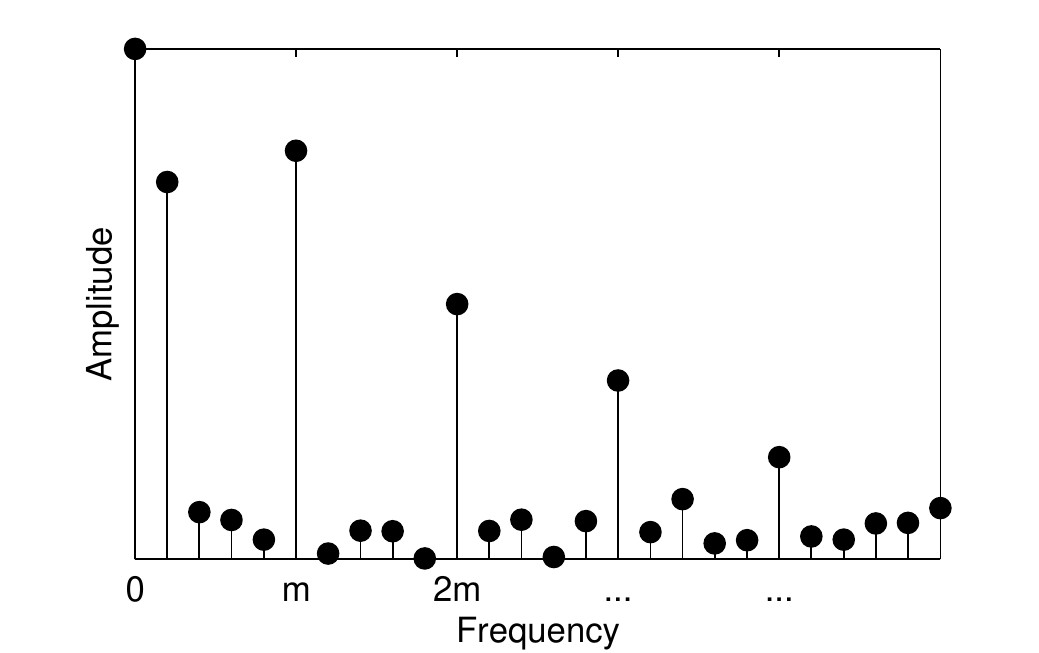}
\caption{Spectrum of $F(k)$ generated from tile type III}
\label{spec5}

\end{figure}

In the area of signal processing, padding zero value to the end of the signal could enhance the resolution of the spectrum. In other words, the number of points used to observe the spectrum increases. Comparatively in this case, after FFT, the interval between two consecutive main components will no longer equal to the height ($n$) of the original image. For instance, the main components are located at $kn'$ in Figure~\ref{spec4} instead of $kn$ in Figure~\ref{spec3} even though the size of two images are the same. On the other hand, theoretically, the number of the main components is not changed, which can be used to infer the width ($m$) of the image. Yet, this approach is not robust when the main components are not prominent (e.g., the spectrum of component at $(m - 1)*n'$ is close to the neighboring components in Figure~\ref{spec4}). \textbf{We solve this problem through another round of FFT over $F(k)$ and pick the main component with the highest spectrum from the result, denoted $F_{F(k)}$.} This approach works because the main components occur every $n'$ points, which suggests its occurring frequency is $m$. Figure~\ref{spec5} illustrates the spectrum of Figure~\ref{spec4}. Clearly, the component with the maximum amplitude is located at $m$ after filtering out low frequency components by HPF (High Pass Filter). While the main components at $km$ ($k > 1$) also show much higher amplitudes than neighboring components, their amplitudes are lower than the amplitude of component at $m$. This can be explained from the nature of image: the similarity between a pair of interleaving rows should be relatively high but still lower than that between a pair of consecutive rows. After $m$ is computed, we can derive $n$ by dividing $m$ from $N$. $n$ might be inaccurate since tile in this case contains more elements than the image object, which will be adjusted in the padding removal step.

Prior to picking out main components from $F_{F(k)}$, we remove the low-frequency components first (achieve HPF). The low-frequency components could have high spectrum amplitude (e.g., when frequency is $0$ its values is smaller than the value at $m$), because neighboring pixels are all similar, which results in a lot of low-frequency components. We use a threshold here to prune the low-frequency components. A threshold too high (more than $m$) would directly filter out the right answer. If it is too low, a wrong main component would be selected. We set it to the first frequency point whose amplitude below the mean value, because we found the low frequency components decreases very fast. Their values will decrease below the mean value within some points while the first main component is much higher than mean.

\vspace{2pt}\noindent{\bf Non-zero Paddings.} We assume the paddings ahead and after the image are all 0 in the last simplified case which does not hold for all tiles extracted from real-world GPU dump. They could be filled by any value. The spectrum would be changed when they happen to show different periodicity and are the tile is not long enough. The chances, however, are low, given the quantity of meaningful pixels are usually much more than the paddings. \ignore{So, the non-zero paddings nearly don't impact our analysis. And our evaluation result supports this hypothesis.} Therefore, the approach for tile type III can be applied to calculate $m$ for this case (both leading and trailing paddings are non-zero) without any modification.

\begin{algorithm}[ht]
  \caption{ImageRecover}
  \label{algorithm}
  \begin{algorithmic}
    \State{Input: $f$;} \Comment{$f$ is the tile}
    \State{$F_1 \gets abs(fft(f))$;}
    \State{$F_2 \gets abs(fft(F_1))$;}
    \State{$F_m \gets mean(F_2)$;}
    \State{$F_2 \gets F_2 / F_m$} \Comment{Normalization}
    \State{$cutFreq \gets locateFirstSmallerThanOne(F_2)$;}
    \State{$F_2(0:cutFreq-1) \gets 0$;} \Comment{High Pass Filter}
    \State{$m \gets locateMax(F_2)$;}
    \If{$ F_2(m) < \theta_0$}
        \State{$Throw(notEnoughLength)$}
    \EndIf
    \State{$N \gets length(f)$;}
    \State{$n \gets \lfloor N/m \rfloor$;}
    \State{$a \gets reshape(f(0:n*m-1),n,m)$;}
    \State{}
    \For{$i \in [0:m-1]$}
        \State{$dist[i] \gets distance(a(:,i),a(:,(i+m-1) mod~m))$}
    \EndFor
    \State{$dist \gets dist/mean(dist)$;} \Comment{Normalization}
    \State{$s \gets locateMax(dist)$;}
    \State{$s' \gets locateSecond(dist)$;}
    \If{$dist(0) < \theta_1~\&\&~dist(s) / dist(s') > \theta_2 $}
        \State{$n \gets \lfloor (N - s)/m \rfloor$;}
        \State{$a \gets reshape(f(s:s+n*m-1),n,m)$;}
    \EndIf
    \State{Output: $a$;}
  \end{algorithmic}
\end{algorithm}

\vspace{2pt}\noindent{\bf Padding removal.} 
Finally, we propose ways to compute $s$ and remove the leading block. Without removing the leading block, the recovered image will not be correctly aligned, such an example is shown in Figure~\ref{all_1}. We elaborate how to derive $s$ for reshaping the image (see Figure~\ref{all_2}) below.

Our computation is again based on the observation that the consecutive columns should be sufficiently similar. Imagine the elements of a tile are placed sequentially into a matrix after $m$ is correctly inferred (see Figure~\ref{all_1}). To transform this matrix to the original image matrix, each row of the matrix should be shifted left for $s~mod~m$ elements if $s$ elements are posited ahead. We aims to calculate $s~mod~m$ and remove those paddings.

The first and the last columns of the tile matrix should be quite similar as they are in fact $(m-s)_{th}$ and $(m-s-1)_{th}$ columns in the original image. On the other hand, the $s_{th}$ and $(s-1)_{th}$ columns of the tile matrix should be quite different as they are in fact the first and last columns (or boundaries) in the original image.

We leverage the findings above to design the following algorithm to infer $s$. First, we build a distance array $dist$, where the $i$th element stores the distance between $i_{th}$ and $(i-1)_{th}$ columns ($\forall i \in [1, m-1]$) of tile matrix, and $dist[0]$ stores the distance between the last and the first columns. The distance between two columns is calculated as counting the number of element pairs of which the differences are larger than a predefined threshold $\theta_3$. If $dist[0] / mean(dist) < \theta_1$ and $max(dist) / second(dist) > \theta_2$ ($\theta_1$ and $\theta_2$ are two thresholds), there exists $s$ leading elements and $s$ is set to be the index of the maximum element in $dist$. If the first check using $\theta_1$ is satisfied, the first and last column are pretty similar and they should be located in the middle of original image. If the check with $\theta_2$ is satisfied, it implies a column pair in the middle of the matrix somewhere has a distinctively low similarity and should be the real image boundary. Then, we remove the first $s$ elements from the tile and reshapes the tile to a matrix with width $m$ and height $(N-s)/m$. Figure~\ref{all_2} illustrates the final image.

To notice, we do not attempt to infer the position of trailing block and remove it, because the trailing block only brings in additional lines below the image when displayed. Likewise, when $s>m$, our algorithm removes $s~mod~m$ elements and leaves additional lines above the image. These additional lines would not prohibit the adversary from recognizing the texts and objects.

\vspace{2pt}\noindent{\bf The algorithm and parameters.} The whole algorithm including preprocessing, identifying the number of rows and columns, and removing leading block is shown in Algorithm~\ref{algorithm}. We found if the input tile $f$ is not long enough, the main components may be overwhelmed by other components. So, we make a parameter $\theta_0$ and set it to 1.5 in our evaluation. It is used to warn attackers when the main component we found is not high enough. When the first main component we found is less than the threshold, it is likely that the tile inferred is incorrect and we mark it with the ``potential false-positive'' before sending out to the attacker. The other 3 thresholds $\theta_1$, $\theta_2$ and $\theta_3$ are set to 2, 1.2 and 5 respectively after parameter tuning by preliminary tests.

%% file: evaluate.tex
\begin{table*}[t!]
\centering
\scalebox{0.9}{
\begin{tabular}{ l c c }
    &AMD Platform &Nvidia Platform\\ \hline
  GPU & HD 6350 (CEDAR) / Sapphire R7 250X & GTX 750 (Maxwell GM107)\\
  Video Memory & 512MB / 1GB & 1GB \\
  GPU Driver Version& fglrx 15.200 & 340.29 \\
  OS Version& Ubuntu 14.04 LTS & Ubuntu 14.04 LTS \\
  CPU & Intel Xeon E3-1225 v2 & Intel Core 2 Duo E8400 \\
  Main Memory & 24GB & 4GB \\
\end{tabular}
}
\caption{Platforms used for evaluation.}
\label{env}
\end{table*}

\section{Evaluation}
\label{eval}


During the evaluation, we first evaluated the accuracy of the recovery algorithm in both single machine and virtual machine, which suggests that our algorithm works well. To more clearly demonstrate its impact to real world, we also evaluate our image recovery attack against popular desktop applications, which are extensively used for image or text rendering. The result is surprising: not only do we show the attack can succeed in totally different applications, we also recover users' sensitive information like account name, email titles. Compared to previous research (e.g., Lee et. al. ~\cite{LeeKKK14}), the attack surface is broader and the information revealed is far more substantial. We elaborate the settings and results as follow.

\subsection{Testing Environment and Performance}


We conducted the evaluation on AMD and Nvidia platforms. The specifications of testing environment are described in Table~\ref{env}. Malicious application we developed uses OpenCL APIs to operate GPU memory and Matlab functions to recover image. We demonstrate the effectiveness of our attack against 4 popular applications on Ubuntu: Google Chrome, Adobe PDF Reader, GIMP, Matlab. Except Matlab, all the other applications are run on AMD platform as OpenCL is natively supported. Matlab is evaluated on Nvidia platform as it requires CUDA support to operate GPU, which is only available on Nvidia platform. Though there is a 3rd-party toolkit named opencl-tool-box that enables Matlab developers to use GPU resources on AMD platform which only supports OpenCL, it is not incorporated into Matlab's official release and has not been updated since Jan 2013 \cite{ocltool}. Therefore, we did not test Matlab on AMD platform.

Our attack against the applications follows the same routine: the malicious application we built initializes the GPU memory and monitors the usage of GPU memory. Then, the victim application is launched and we simulate a series of users operations, like viewing a web page and viewing a PDF document. Finally, the victim application is closed and the malicious application is reactivated due to the sudden increase of available memory. The GPU memory is instantly dumped and analyzed by the malicious application for image recovering. Finally, the malicious application saves the restored image as image files formatted in either RGBA or ARGB which is decided during the step of data blocks pruning. To notice, we did not make GPU memory exclusive to the malicious and victim applications during experiments.


The overhead of each attack is bounded to the specifications of platform and the layout of GPU but it is in general unnoticeable. We run our malicious application against each victim application for 5 times  and calculate the average time consumed in different steps. It takes 75 to 95 ms for memory initialization and 110 to 130 ms for data blocks extraction \& pruning on AMD platform. While on Nvidia platform, the overhead significantly increases. It takes 350 ms for memory initialization and 550 ms for data blocks extraction \& pruning. We speculate the overhead increases mainly due to the larger memory of the Nvidia platform. The overhead for layout inference is bounded to the size of tile (the time complexity of Algorithm~\ref{algorithm} is $O(n\log n)$ where $n$ is the tile size). The largest tile we encountered is 15MB and can be processed in 13ms. Meanwhile, the number of tiles for a memory dump is up to 625 among all the experiments. The overhead of this step in most cases would not exceed a second. In total, the attack could end in several seconds which hardly raises the suspicion from user. The highest CPU usage we observed during the inference phase is 45\%.

Our accuracy test result shows that almost every image can be recovered in different size, brightness, noise etc. The detailed result is included in the Appendix. Next, we describe our attack result against top-tier victim applications in details.

\subsection{Accuracy Evaluation}

Before testing against real-world applications, we evaluated the accuracy of our approach in reconstructing the original image. To this end, we developed a toy application whose sole task is to load an image into GPU memory. In particular, the application reads one JPG file from the set of test JPG files on disk, decodes it into bitmap format, stores it in GPU memory and then exits without zeroing out the used memory region. The malicious application will then attempt to reconstruct the original image from the uninitialized memory. The test ends when all JPG files are loaded. We did not use other commercial or open-source applications since they may split the image into pieces and render them in parallel.

The testing image set comes from the 29 sample images from INRIA Holidays dataset, which is widely used for evaluating computer vision algorithms \cite{jegou2008hamming,imagSet}. We begin with the evaluation on the accuracy of recovering original sample images. Next, we zoom out those images to different sizes and assess the impact of image size to our approach. At last, we apply different types of transformation on the sample images to understand the limitation of our approach, i.e., which factor impedes the success reconstruction by our approach.

\begin{table}[h]
\centering
\scalebox{1}{
\begin{tabular}{ c | c  c c}
  Scale &  Typical   &   Successfully  &  Recovered but \\
  Ratio &  Size      &   Recovered     &  not in Samples        \\ \hline
  1     &  1024*768  &   29            &  18            \\ \hline
  0.5   &  512*384   &   29            &  8             \\ \hline
  0.25  &  256*192   &   29            &  7             \\ \hline
  0.125 &  128*96    &   29            &  13            \\ \hline
  0.0625&  64*48     &   29            &  28            \\
\end{tabular}}
\caption{Accuracy test for the self-developed application.}
\label{AccuTestResult}
\end{table}

Table~\ref{AccuTestResult} shows the test result for the initial two evaluation tasks. Specifically, all of the original images are successfully recovered. When scaling the size ratio of the image from 1 to 0.0625 (the size is down to 64*48, the icon size), the result is not changed with all images successfully restored, which indicates our approach is robust against images with varying sizes. Interestingly, we also recovered images which were not loaded by our application sometimes. We suspect these images were rendered by applications running simultaneously with our application.


Then we test the capability of our approach in dealing with less meaningful images. Adding noise is a common way to obscure the meaningful pieces within images and we apply Gaussian noises on the sample images and check whether they can still be restored. We add Gaussian random number falling within the range of $N(0,\sigma)$ ($\sigma$ is the noise standard deviation and the larger $\sigma$ means more noisy) to each pixel of the original images before loading them into GPU memory. Table~\ref{NoiseTestResult} shows the number of successfully recovered images under different settings of $\sigma$.

\begin{table}[h]
\centering
\scalebox{1}{
\begin{tabular}{ c | c c c c c c}
Noise $\sigma$ & 1  & 5  & 10 & 20 & 30  & 40 \\ \hline
Recovered \#   & 29 & 29 & 28 & 26 & 25  & 23
\end{tabular}}
\caption{Successfully recovered images under different noise settings.}
\label{NoiseTestResult}
\end{table}

\begin{figure}[h]
    \center
     \begin{subfigure}[h]{0.22\textwidth}
        \includegraphics[width=\textwidth]{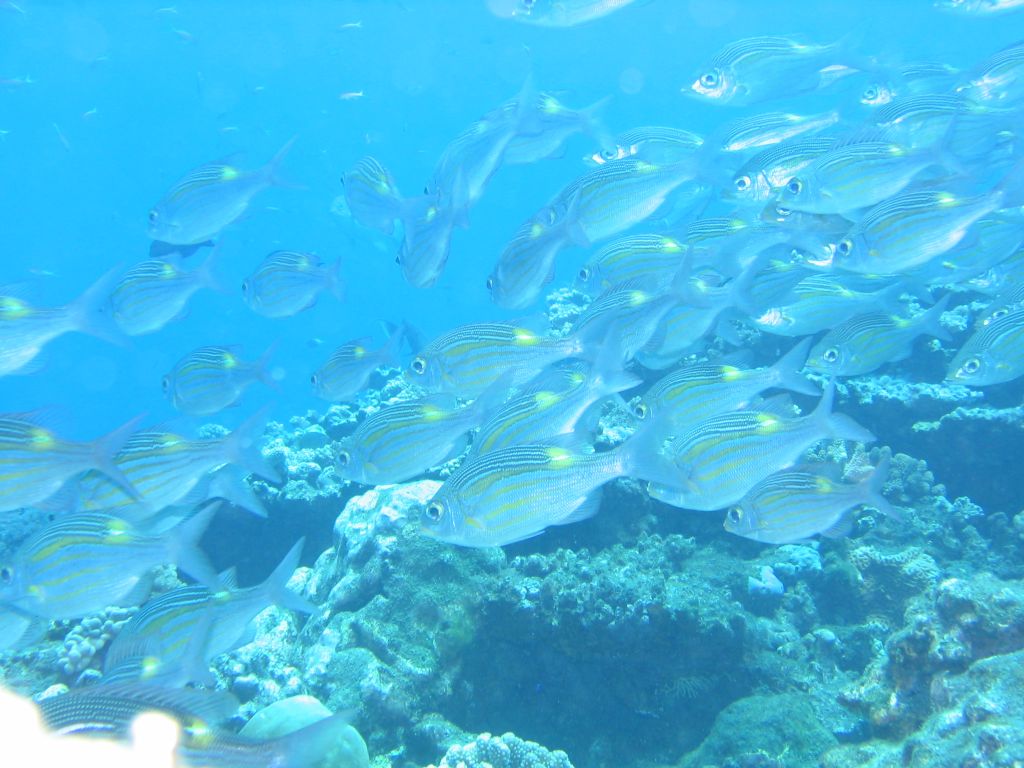}
        \caption{Original image.\\\quad }
        \label{fig:NoiseOriginal}
     \end{subfigure}
     \begin{subfigure}[h]{0.22\textwidth}
        \includegraphics[width=\textwidth]{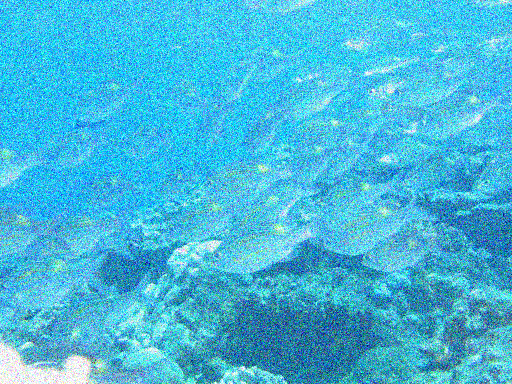}
        \caption{Image interfered with noise.}
        \label{fig:Noised}
     \end{subfigure}

     \begin{subfigure}[h]{0.22\textwidth}
        \includegraphics[width=\textwidth]{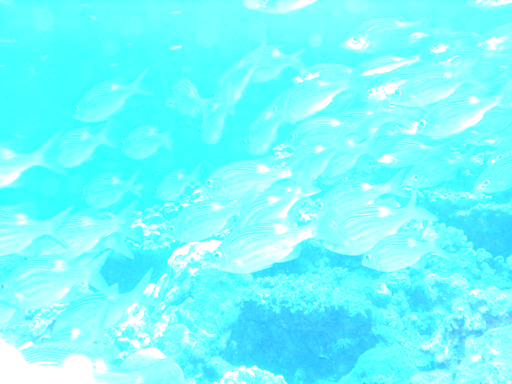}
        \caption{Image with increased brightness.}
        \label{fig:Brighter}
     \end{subfigure}
     \begin{subfigure}[h]{0.22\textwidth}
        \includegraphics[width=\textwidth]{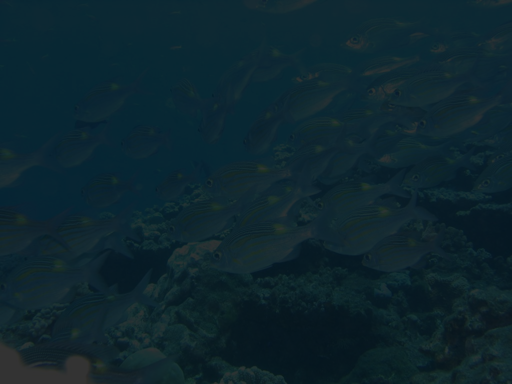}
        \caption{Image with decreased brightness.}
        \label{fig:Brightless}
     \end{subfigure}

     \caption{The original image and the transformed versions.}
     \label{fig:NoiseComp}
\end{figure}

As suggested by the result, our algorithm can recover images even when they are interfered by large Gaussian noise. When the noise standard deviation $\sigma$ is increased to 40, the interfered images are barely recognizable to human, yet they can still be recovered with high success rate at 79.3\%. Figure~\ref{fig:NoiseOriginal} and Figure~\ref{fig:Noised} show the the original image and the image interfered by Gaussian noise when $\sigma=40$. Both images can be correctly restored from memory dumps.

We also assess the impact from other image transformations, including adjusting brightness and contrast. We increased and decreased the brightness and contrast of the 29 images by 80\% separately, and the resulting images are already unrecognizable. Figure~\ref{fig:Brighter} and Figure~\ref{fig:Brightless} show the images after brightness is adjusted\footnote{The images with new contrast setting are not shown here as they are even less discernible.}. Surprisingly, all such images are restored by our approach with 100\% success rate. The result strongly supports the robustness of our approach.

\subsection{Virtualized Environment Experiments}
With the advent of cloud, there are increasingly more companies start to rent virtual machines running on their spare computing platforms to users with computation demands but don't want to manage physical machines. To satisfy users with strong computation demand, service provider provides optional GPU support to their virtual machines. And users can rent one virtual machine with GPU to run their GPU-accelerated programs. We want to know whether our method can be applied to such a virtualized environment because it implies abundant victim users.

We have rented a GPU passthrough capable GPU, the Sapphire R7 250X, to set up a virtualized test bed. The GPU of the AMD platform was temporally replaced by the rented card to make virtual machine GPU capable. We used QEMU 2.4.50 as the hypervisor to run virtual machines with GPU passthrough.

Our evaluation procedure follows this routine: attacker's VM is started first to initialize the GPU memory and then shut down. Next, victim turns on his VM, uses our self-developed application used for accuracy evaluation to load the 29 images to GPU memory and then shut down. At last, attacker starts his VM to extract GPU memory and recover images. One may wonder why we do not run two VMs at the same time. That's because a GPU can be passed through to exactly only one VM. During the VM switching process, the physical machine cannot be restarted because restarting the physical machine will reset the GPU and its memory.

Our evaluation results show that, among the 29 images, 25 was completely recovered while 2 was completely missing and the left 2 images can be recovered partially. The result is not as good as what in single machine context. We believe it is because the time gap between the termination of victim app and residues extraction is increased. When attacker has a process running together with victim, he can monitor the GPU memory usage and extract residues immediately after the victim application terminates. However, in the virtualized context, attacker can only extract the residue at least after a VM switching process, which leads to a lot of uncontrollable factors that may pollute the memory. But, the ratio of recovered images is still prominent, indicating the threat to virtualized environment cannot be neglected.
\subsection{Case 1: Google Chrome}
\label{chrome}
More and more web applications are developed to process users' personal information nowadays. Browser vendors designed various mechanisms to protect users' data, like private browsing. These mechanisms intend to defend against malicious web pages or extensions planted by attackers but are powerless against the adversary capable of stealing GPU memory. The problem exacerbates in up-to-date browsers where GPU-acceleration is intensively used. We use Google Chrome as an example to demonstrate the seriousness of this problem. Gmail is used here as a showcase to demonstrate what types of content can be recovered by our attack. In addition, we exercise automated information extraction techniques against memory dumps from different web sites to assess the overall impact of the attack.



\vspace{2pt}\noindent{\bf Recovered content from Gmail.}
In this attack, we assume the victim user logs into her Gmail account and the email titles are all displayed. We run the analysis routine against the page of email list of one user and are able to recover 113 images from the GPU memory dump, among which, the largest one has 512K pixels, the smallest has 926 pixels and the average size of the images is 23.74 KB (PNG format). The images are manually classified upon their visual positions in the browser UI. The details of the leaked images are described below separately:

\begin{figure}[ht]
\centering
\includegraphics[width=0.3\textwidth]{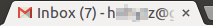}
\caption{Tab of Gmail}
\label{gtab}
\end{figure}

\vspace{3pt}\noindent$\bullet$\textit{  Tab:} 
The tab of Google Chrome displays the favicon and the title specified by the web page. It could tell which web site is visited by the user. Moreover for Gmail page, the information revealed is more than just the name of web site. As shown in Figure~\ref{gtab}, the email address and the number of unread emails are also displayed in the tab. Leaking email address is of course unwanted for the victim as it can be exploited to send targeted phishing emails or harvest user's social profiles by querying popular social network sites.  What's worse, this issue is not unique to Gmail and equal or more information could be disclosed from the tab of other sites. For instance, Amazon displays the name of the product user is viewing and YouTube displays the name of the video user is watching.

\begin{table*}[t!]
\centering
\scalebox{1}{
\begin{tabular}{ c c c c c c }
    & www.gmail.com & www.youtube.com & www.yahoo.com & www.facebook.com & www.twitter.com \\ \hline
  Characters & 2388 &9184&690&6183&2110\\
  Words & 416 & 453&215&826&481 \\
  Faces& 0 & 24&8&8&16 \\
\end{tabular}}
\caption{Leakage from different websites}
\label{browser}
\end{table*}


Though our initial exploration indicates that critical information could be leaked, the extent of the inferred result should be taken a grain of salt. Google Chrome limits the number of characters displayed in a tab. The tab will be squeezed when many tabs are opened (it begins to resize when more than 7 tabs are opened in a 14-inch laptop with screen resolution set to 1600x1200). This design results in partial reveal of the title of a web site: for example, the Gmail tab only displays the first 14 characters of user name (under the same screen setting) and therefore a long user name will not be fully recovered. However, a large number of users choose short user names\cite{username} and knowing 14 characters is still a big lift if the adversary plans brute-force crack.

\begin{figure}[ht]
\centering
\includegraphics[width=0.272\textwidth]{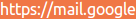}

\includegraphics[width=0.316\textwidth]{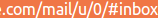}
\caption{Address bar of Gmail}
\label{gadd}
\end{figure}

\vspace{3pt}\noindent$\bullet$\textit{  Address bar:} 
An image containing the address bar is also recovered by our program and is shown in Figure~\ref{gadd}. To notice, the image does not show the whole region of the address bar or even the full URL. In fact, Chrome attempts to render the address with GPU when ``AutoComplete'' is turned on and user is typing. The recovered image reflects the characters that have been input. Though not fully displayed, this partial image can still tell that the user is using Gmail. When the user is not typing in address bar, a different type of image will be generated and an example is shown in Figure~\ref{Addr}. By collecting the leaked information from such images (also combining with tab images), an adversary is able to partially reconstruct user's browsing history which clearly violates user's privacy. Previous research by Lee et. al. ~\cite{LeeKKK14} profiled 1000 web site homepages ahead and attempts to identify which one has been visited. Our attack takes a big step forward as potentially any site visited can be inferred without prior profiling and it is also resilient to the content change of web sites.

\begin{figure}[ht!]
\centering
\includegraphics[width=0.3\textwidth]{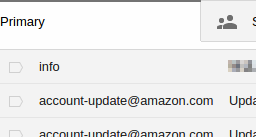}
\caption{Part of Gmail inbox}
\label{glayer}
\end{figure}

\vspace{3pt}\noindent$\bullet$\textit{  Page body:} 
Most of the images recovered can be attributed to the page body. Figure~\ref{glayer} shows one segment of Gmail inbox content and the senders and initial characters of emails are displayed which are obviously sensitive to the user. \ignore{What information is leaked depends on the web sites visited but clearly this poses a big threat to user. Any sensitive information in the page is within the reach of adversary.} Since current web applications are designed to intensively deal with personal information, the threat could be more substantial if the malicious program is able to run on the victim's machine for long time and restore images from different web pages. Among these recovered images, we also found seemingly unmeaningful textures like the border of an object box. We suspect they are peeled from the original objects due to browser's splitting algorithm. They can be combined with their counterparts through puzzle-solving algorithm.


\vspace{2pt}\noindent{\bf Automated information extraction.}
All the images are manually examined for this Gmail case but it is not scalable, especially when a large number of users are monitored or the tabs of Chrome are frequently closed and opened. We want to reduce attacker's workload by only sending the sensitive images for analysis. In fact, it is quite challenging to automatically select the sensitive images, which requires extensive knowledge of user's background and application's context. A more practical goal could be identifying the images enclosing texts and faces, which are already meaningful in common scenarios and can be automated.

For this purpose, we use Adobe PDF professional OCR module to find texts (the images need to be converted into PDF first) and build a Matlab program leveraging a widely-used library computer vision system toolbox to recognize faces. Only the images containing either texts or faces are passed to the next step, i.e., analyzed by the attacker. We evaluate them on Gmail images and reduce the number of images of interest to 31 (out of 113 images in total) while all images about tab, address bar and inbox are identified. The overhead incurred in this process is also small, only costing several seconds in OCR and face recognition. We assume the modules are run on attacker's server but they could be run on victim's machines as well. For the latter setting, only detected images are transmitted, which significantly reduces the network overhead and makes the attack even stealthier.

\begin{figure}[ht!]
\centering
\includegraphics[width=0.25\textwidth]{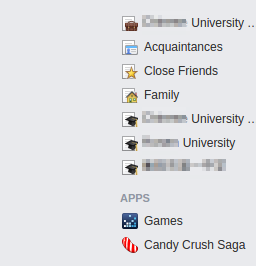}
\caption{The list of attended universities and schools identified from user's Facebook page}
\label{facebook}
\end{figure}

Our attack against Gmail shows sensitive information can be successfully revealed but it is unclear yet whether the issue is universal. We try to answer this question by evaluating our algorithm on 4 other web sites with top popularity: YouTube, Yahoo, Facebook, Twitter. Since there is no existing metric to quantify the sensitive data leaked, we choose to measure the number of words and faces recognized from the images. As shown in Table~\ref{browser}, approximately hundreds of words and tens of faces could be identified for each web site. Though not all texts and faces are sensitive (e.g., we found a lot of them are related to advertisements), the chances of leakage are still high. Particularly, the social network sites like Facebook and Twitter tend to leak more useful information. Tweets and Facebook posts have been discovered among the recovered images. Figure.\ref{facebook} shows a tile exhibiting Facebook user's educational experiences (sensitive personal information is mosaicked). Besides, we also tried to evaluate our attack against e-banking. The results showed to attacker the \textbf{account balance}, \textbf{credit card account number} and \textbf{transaction details}.

\vspace{2pt}\noindent{\bf Discussion.} Finally, we want to understand why segments are extracted rather than the whole web page. According to the design document\cite{chroGPU}, Chrome breaks the page into small tiles and allocates GPU resources for some of them based on their predefined priorities\cite{TileDesign}. Besides, not all image segments are preserved in GPU memory when dumped by our attack program. Under the real-world settings, there are GPU memory restrictions that limit the number of tiles residing in GPU and memory manager is allowed to evict tiles from GPU memory. Therefore, not all segments can be recovered.

We not only demonstrate our attack on Google Chrome but Firefox is also under threat. In fact, it is confirmed that Firefox also leaves residues in GPU memory~\cite{LeeKKK14}. We tested gmail against Firefox. Firefox does not produce residue images related to address bar and tab caption, but more severely, \textbf{it yields a relatively complete block showing the page body}. In gmail case it is a large image containing the sender, title and part of email contents. Such leaked information is definitely more valuable to attackers.

\subsection{Case 2: Adobe Reader}
Adobe Reader uses GPU to accelerate the rendering process of PDF documents. We consider the texts and graphs of a PDF as sensitive and tests if they can be extracted by exploiting the residues left by the application. We use a PDF of a research paper as an example and the content recovered is described below.


\vspace{2pt}\noindent{\bf Recovered content from PDF.} It turns out both the graphs and texts are rendered in GPU. Similar to the Chrome case, segments of figures and texts are recovered. Interestingly, the segments do not only belong to the page shown at foreground, but some of them also belong to the pages rendered at background.

\vspace{3pt}\noindent$\bullet$\textit{  Fragments of figures:}
We found some of the recovered images are actually fragments of a given figure. We have not tried to combine the fragments to recover the original figure but it is possible certain algorithm could achieve this goal, for example, by taking the advantage of similarities among the edges of neighboring fragments.

\begin{figure}[ht]
\center
\begin{subfigure}[b]{0.5\textwidth}
\center
\includegraphics[width=0.8\textwidth]{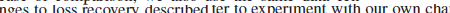}
\caption{A normal line of text}
\label{ar1}
\end{subfigure}

\begin{subfigure}[b]{0.5\textwidth}
\center
\includegraphics[width=0.8\textwidth]{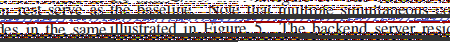}
\caption{A line of text with noise}
\label{ar2}
\end{subfigure}
\caption{Two typical images recovered from residues left by Adobe Reader}
\end{figure}

\vspace{3pt}\noindent$\bullet$\textit{  Lines of text:} 
It turns out Adobe Reader separates text region into long stripes. Figure~\ref{ar1} shows a normal line of text recovered (some letters are incomplete) and Figure~\ref{ar2} shows one line of text with noise above and below. Though not fully recovered, the text from the images can be easily read by attacker. When too many images are extracted from GPU, the adversary can use the OCR and natural language processing technique to reduce the number of images requiring manual analysis.


\subsection{Case 3: GIMP}
In this section, we evaluate a popular image processing software on Unix platform, GIMP. Under the default configuration, GIMP does not rely on GPU to render images. However, when the image is large, the user is recommended to turn on GPU acceleration, which can be done through linking to a graphics library named \texttt{GEGL} when GIMP is started. A user can simply pass \texttt{"GEGL\_USE\_OPENCL=yes"} when launching GIMP. Our attack is tested under this setting. Specifically, we opened an image file, applied some different image filter (e.g., edge-laplace) for each running, and then closed the image file.

\vspace{2pt}\noindent{\bf Recovered content from GIMP figures.}
Our evaluations showed the type of filter will decide the outcome of the attack. For some filters, nothing can be revealed from the image either before or after applying the filters. But for other filters, the recovered images are actually close to the \textbf{whole} original images having been passed to GIMP, without any fragmentation.


\begin{figure}[h]
\center
\begin{subfigure}[b]{0.3\textwidth}
    \center
    \includegraphics[width=\textwidth]{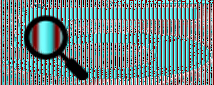}
    \caption{Before}
    \label{stripe}
\end{subfigure}
\quad
\begin{subfigure}[b]{0.0745\textwidth}
    \center
    \includegraphics[width=\textwidth]{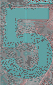}
    \caption{After}
    \label{After}
\end{subfigure}
\caption{Images recovered from leakage with stripes before and after the compression}
\end{figure}

\ignore{\textbf{Perfectly Recovered}
For some filters, we can recover images after filtering flawlessly. Similar to previous conditions, the image quality doesn't degrade.}

Besides, we also restored images with vertical stripes intersected with the original images (see Figure~\ref{stripe}). This case is likely caused by the implementation of GEGL which does not adopt the standard image format (RGBA and ARGB). Since such stripes could affect the step of information inferring, we remove them with a ``compressing'' process: the width of the recovered image is shrunk to 25\% and a pixel in the new image is combined from the 4 consecutive pixels horizontally. The image derived from Figure~\ref{stripe} is shown in Figure~\ref{After}.

This case suggests that even if the developer does not use standard RGBA or ARGB format, the sensitive information of the image can still be extracted, by tweaking our approach. The pseudo-periodicity still holds despite the loss of the color map information, which results in produced images with correct alignment but inaccurate color. However, a large portion of the image can be recognized with techniques like text recognition.



\subsection{Case 4: Matlab}

Matlab is widely used by academia and industry for scientific computing. It also provides various libraries to support image processing, and developers could leverage the power of GPU with parallel computing toolbox. We assume the developer first loads from hard disk drive a picture and then converts it into a GPU-compatible object ``gpuArray''. This image object will be passed to GPU for processing, and Matlab is closed when processing finished. Finally, the residues in GPU memory are dumped and analyzed.


\vspace{2pt}\noindent{\bf Recovered content.}
The loaded picture is split by Matlab into fragments and therefore the whole picture can not be directly recovered. However, the size of the fragments is still large enough which makes it possible to partially or fully restore the original image with rearrangement. After the fragments are put together, we found the generated picture is in fact flipped along the diagonal of original picture. This is because Matlab stores a picture into a matrix through column-by-column instead of normal row-by-row fashion, making the image automatically transposed in memory. Thus our proposed algorithm can still be applied by simply transposing the image matrix back.

\vspace{2pt}\noindent{\bf Discussion.}
Matlab is a very popular computing tool, which has already been used for many different fields. Matlab with Parallel Computing Toolbox is often run by developers on high-performance computer with powerful GPU. The high-performance computer is shared by different users in many cases, since the computing resources are expensive and it is a waste when the computer is idle. So, there is a higher chance where the attacker can get victim users' images from these kinds of computers.

%% file: end.tex
\section{Related Works}
\label{relat}

\vspace{2pt}\noindent{\bf GPU vulnerabilities.} As the techniques for GPU computing advance, different security issues also emerge. Previous research shows the defense around GPU is far from perfect \cite{patterson2013vulnerability,jeonarchitectural}. Lombardi et. al. carried out a comprehensive analysis over GPU used in cloud and revealed several leakage issues. Pietro etc. discovered leakage in CUDA framework and their evaluation showed that global memory, shared memory and registers are all vulnerable ~\cite{di2013cuda}. Cl{\'e}mentine et. al. proposed an attack to acquire leakage across virtual machines ~\cite{maurice2014confidentiality}. In a work by Ladakis et. al.~\cite{ladakis2013you}, they implemented a stealthy keylogger using GPU. The closest work to our research is done by Lee et. al.~\cite{LeeKKK14} in which they are able to infer which web site has been visited by a victim based on color distribution of GPU memory. Instead, we are able to recover images with sensitive information.

\ignore{\vspace{2pt}\noindent{\bf CPU cache vulnerabilities.} The security of CPU cache were comprehensively studied by previous researchers \cite{osvik2006cache, tromer2010efficient, percival2005cache, page2002theoretical, bonneau2006cache}. Cache is also shared among different processes, by exploiting timing or other patterns, a malicious process can steal the secret of another process, including the encryption keys. Particularly, Ac{\i}i{\c{c}}mez et. al. proposed a scheme to obtain keys of remote AES cryptosystems based on cache timing\cite{aciiccmez2006cache}. They also proposed a scheme to attack OpenSSL in \cite{aciiccmez2008vulnerability} later. On the defense side, Keramidas et. al. proposed a non-deterministic cache scheme against side cache attacks~\cite{keramidas2008non}. There are also some other defense schemes proposed by Page et. al.\cite{page2003defending,page2005partitioned}. Our research shows the GPU memory is also vulnerable and the information leaked is much more than just pieces from side channels.}

\vspace{2pt}\noindent{\bf Memory forensic.} \ignore{There are a lot of recent works concerning memory forensic. Saltaformaggio etc. proposed a method to re-construct Android APP's GUI displays by re-constructing the GUI tree topology and reconstruct the drawing operations~\cite{saltaformaggio2015guitar}. He also introduced a method to recover photographic evidence produced by smartphone camera by exploiting the memory possessed by the intermediary service ~\cite{saltaformaggio2015vcr}. He also showed a method to re-use application's logic to recover images from criminal's phone memory~\cite{saltaformaggio2014dscrete}.} Memory forensic has been studied long time ago as a way to help government and police collect electronic evidences from criminal's devices. In recent years, development has been made in recovering images from main memory for forensic needs. Saltaformaggio et. al. proposed a method to reconstruct Android APP's GUI displays by reconstructing the GUI tree topology and reconstruct the drawing operations~\cite{saltaformaggio2015guitar}. They also introduced a method to recover photographic evidence produced by smartphone camera by using the memory possessed by the intermediary service ~\cite{saltaformaggio2015vcr}. In addition, a method to re-use application's logic to recover images from criminal's computer memory~\cite{saltaformaggio2014dscrete} was proposed recently. These works assume pre-knowledge of the applications has been acquired and apply program analysis for image reconstruction. Under their settings, the data structures for keeping images can be obtained. Our problem is much more challenging as GPU memory keeps no meta-data of images. Still, the image data can be decoded by leveraging the internal similarity feature, as proved by our work.

\section{Conclusion}
\label{concl}
In this paper, we proved GPU memory management strategy is vulnerable, by proposing a novel attack to recover the images from the leftover of other applications or other VMs in GPU memory. Our recovery technique is motivated by the observation that there is strong correlation between rows and columns of an image. By evaluating on highly popular applications, we show the severity of the security problems regarding GPU memory management. Sensitive information like credit card number and email titles can be readily extracted, if it is preciously calculated by GPU. As a result, the severity is underestimated by previous research and the security issues have to be mitigated. 